\documentclass[twocolumn,aps,prb,final,superscriptaddress]{revtex4}

\usepackage{amssymb}
\usepackage{mhchem}
\usepackage{siunitx}
\usepackage{booktabs}

\usepackage[pdftex]{graphicx}
\graphicspath{{figures/}}

\usepackage{color}

\newcommand{\zrte}{ZrTe$_3$}

\newcommand{\fourbar}{(40$\overline{1}$)}
\newcommand{\invA}{\mathrm{\AA}^{-1}}
\renewcommand{\deg}{^\circ}
\renewcommand{\toprule}{\hline}
\renewcommand{\midrule}{\hline}
\renewcommand{\bottomrule}{\hline}
\newcommand{\bbar}{B(A)}
\newcommand{\dbar}{D(E)}

\newcommand{\gammabar}{$\mathrm{\Gamma(Z)}$}

\begin{document}

\title{Local corrugation and persistent charge density wave in ZrTe$_{3}$ with Ni intercalation}

\author{Alex~M.~Ganose}
\affiliation{Department of Chemistry, University College London, 20 Gordon Street, London WC1H 0AJ, UK}
\affiliation{Diamond Light Source, Harwell Campus, Didcot OX11 0DE, UK}
\affiliation{Thomas Young Centre, University College London, Gower Street, London WC1E 6BT, UK}

\author{Liam Gannon}
\affiliation{Diamond Light Source, Harwell Campus, Didcot OX11 0DE, UK}
\affiliation{Clarendon Laboratory, University of Oxford Physics Department, Parks Road, Oxford, OX1 3PU, UK}

\author{Federica Fabrizi}
\affiliation{Diamond Light Source, Harwell Campus, Didcot OX11 0DE, UK}

\author{Hariott Nowell}
\affiliation{Diamond Light Source, Harwell Campus, Didcot OX11 0DE, UK}

\author{Sarah Barnett}
\affiliation{Diamond Light Source, Harwell Campus, Didcot OX11 0DE, UK}

\author{Hechang Lei}
\altaffiliation{Present and permanent address: Department of Physics, Renmin University of China, Beijing, 100872, China}
\affiliation{Condensed Matter Physics and Materials Science Department, Brookhaven National Laboratory Upton, NY 11973, USA}

\author{Xiangde Zhu}
\altaffiliation{Present and permanent address: High Magnetic Field Laboratory, Chinese Academy of Sciences - Hefei 230031, PRC}
\affiliation{Condensed Matter Physics and Materials Science Department, Brookhaven National Laboratory Upton, NY 11973, USA}

\author{Cedomir Petrovic}
\affiliation{Condensed Matter Physics and Materials Science Department, Brookhaven National Laboratory Upton, NY 11973, USA}

\author{David~O.~Scanlon}
\affiliation{Department of Chemistry, University College London, 20 Gordon Street, London WC1H 0AJ, UK}
\affiliation{Diamond Light Source, Harwell Campus, Didcot OX11 0DE, UK}
\affiliation{Thomas Young Centre, University College London, Gower Street, London WC1E 6BT, UK}

\author{Moritz Hoesch}
\affiliation{Diamond Light Source, Harwell Campus, Didcot OX11 0DE, UK}
\email{moritz.hoesch@gmail.com}

\date{\today}

\begin{abstract}
The mechanism of emergent bulk superconductivity in transition metal intercalated ZrTe$_3$ is investigated by studying the effect of Ni doping on the band structure and charge density wave (CDW). The study reports theoretical and experimental results in the range of Ni$_{0.01}$ZrTe$_3$ to Ni$_{0.05}$ZrTe$_3$. In the highest doped samples bulk superconductivity with $T_c<T_{CDW}$ is observed, while $T_{CDW}$ is strongly reduced.  €œRelativistic \textit{ab-initio} calculations reveal Ni incorporation occurs preferentially through intercalation in the van-der-Waals gap. Analysis of the structural and electronic effects of intercalation, indicate buckling of the Te-sheets adjacent to the Ni site akin to a locally stabilised CDW-like lattice distortion.
Experiments by low temperature x-ray diffraction, angle-resolved-photoemission spectroscopy (ARPES) as well as temperature dependent resistivity reveal the nearly unchanged persistence of the CDW into the regime of bulk superconductivity. The CDW gap is found to be unchanged in its extent in momentum space, with the gap size also unchanged or possibly slightly reduced on Ni intercalation. Both experimental observations suggest that superconductivity coexists with the CDW in Ni$_{x}$ZrTe$_3$.
\end{abstract}

\maketitle

\section{Introduction}

The charge density wave (CDW) is a Fermi surface (FS) instability in a crystalline material. Charge carriers self-organise to form a period modulation that can be driven by FS nesting, momentum-dependent electron-phonon coupling or electron correlation effects~\cite{grunerbook,monceau12}. The modulation, characterised by a wave vector $\vec{q}_{CDW}$, can thus be incommensurate to the underlying crystal lattice. It is stabilised by the opening of a gap at the Fermi level $E_F$ and the crystal shows a periodic lattice distortion (PLD). The experimental signatures to look for are thus a resistivity anomaly at the ordering temperature $T_{CDW}$, a gap at $E_F$ detectable in optical or electron spectroscopy and additional scattering below $T_{CDW}$ at wave vectors $\vec{Q}=\vec{G}\pm\vec{q}_{CDW}$,~\cite{monceau12} where $\vec{G}$ is a reciprocal lattice vector. The gap, ideally visible as a band back-folding at the Fermi wave vector $k_F$, is the only signature that accesses the electronic nature of the CDW directly. The question of whether or not the CDW can be stabilised without any interaction with the lattice remains controversial~\cite{johannes08} and thus the combined evidence of all three probes is most suitable to capture the properties of CDW materials.

The removal of Density of States (DOS) at $E_F$ offers a simple mechanism for the CDW to compete with superconducting order (SC), also a FS instability, which requires a metallic state with finite DOS at $E_F$. In many cases the PLD can be tuned~\cite{leroux2015,hoesch2016} and the CDW eventually quenched by the application of hydrostatic pressure, leading to the emergence of superconductivity.~\cite{monceau77,yomo05,hamlin2009,zocco2015} The pressure favours a more three-dimensional electronic structure thus reducing the FS nesting, or in the case of a multi-band system it may lead to a rearrangement of carriers between FS sheets, which can also destabilise the CDW.~\cite{starkovicz07,hoesch2016} Alternatively, disorder may also quench the CDW, which can be achieved by extreme sample growth conditions~\cite{zhu13,denholme2017,li2017hfte3} and in solid-solution samples with isoelectronic substitution.~\cite{zhu2016,li2017}

Intercalation of transition metal atoms into the van-der-Waals gap of the layered transition metal dichalcogenide TiSe$_2$ has attracted particular attention due to the strongly correlated nature of this electron system featuring a commensurate CDW. This is quenched with concomitant emergence of superconductivity by Cu or Fe intercalation.~\cite{morosan06} Significant modifications of the underlying bands have been observed beside the removal of the CDW band folding in the electronic structure when the CDW is quenched.~\cite{cui06,zhao07,Jishi08} The intercalation, as with any doping, introduces disorder. It may lead to a lattice distortion equivalent to (chemical) pressure or the charge transfer from the intercalant can be equivalent to electrostatic doping. A large variety of modified CDW orders has been found in electric field dependent studies of TiSe$_2$.~\cite{li2016} A review of effects of intercalation in layered transition metal chacogenides was given in Ref.~[\onlinecite{chen2016}]. 

\begin{figure}
\centerline{\includegraphics[width = .48\textwidth]{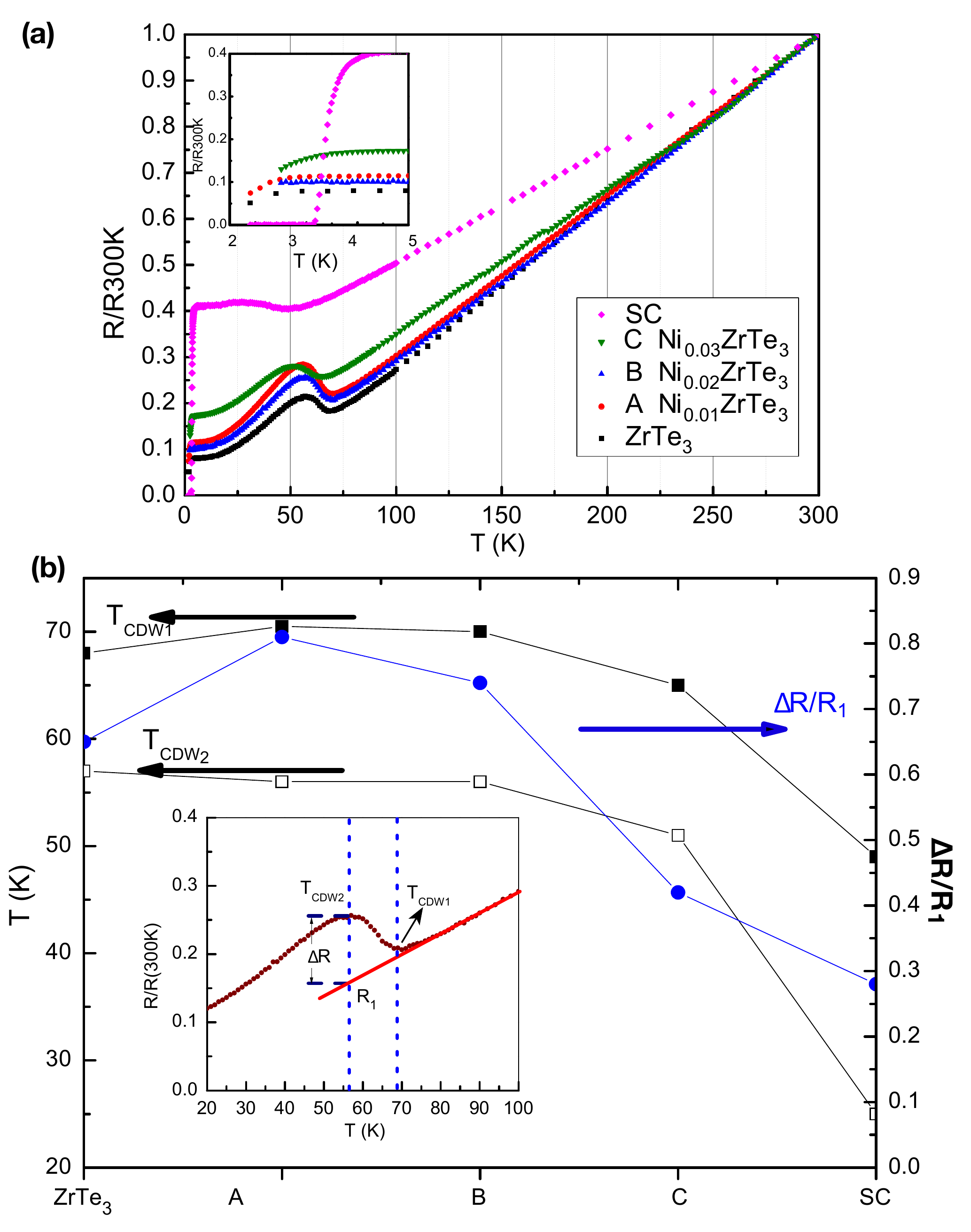}}
\caption{(a) Resistance $R$ normalised to the value at $T=300$~K as a function of temperature for samples A-C and SC as well as pure \zrte (reference). Details of the superconducting transition in the zoomed region from 2 - 5 K are shown in the inset. (b) Extracted Charge Density Wave transition temperatures  $T_{CDW1}$ and  $T_{CDW2}$ for all samples (filled and open squares, left scale) as well as the resistivity anomaly height $\Delta R/R_1$ (circles, right scale). The inset shows details of the definition of $T_{CDW1}$,  $T_{CDW2}$,  $\Delta R$, and $R_1$.}
\label{Fig1}
\end{figure}

\zrte{}, the material of our study, is a metallic member of the family of quasi-one-dimensional transition metal trichalcogenides.~\cite{wilson79} A CDW is seen as a resistivity anomaly around $T_{CDW} = 63$~K~\cite{takahashi83,takahashi1984} with an incommensurate PLD modulation $\vec{q}_{CDW} = (0.071,0,0.336).$~\cite{eaglesham84} The resistivity continues to drop in a metallic fashion below  $T_{CDW}$ and filamentary superconductivity is observed below $T_c \sim 2$~K.~\cite{takahashi83,takahashi1984} Local pairing, evidence of which is found at ambient pressure, may contribute to the stabilisation of bulk SC at high pressures.~\cite{tsuchiya2017} The highly anisotropic crystal lattice consists of prismatic chains (\zrte)$_\infty$, which line up the Zr atoms into chains along $b$. Another chain is formed by Te atoms  linking the chains (Te$^{(2)}$, Te$^{(3)}$ in Fig.~\ref{fig:electronic-structure}), this is where the CDW is formed with a long-wavelength (small $q$) modulation along $a$ and a tripling of the unit cell along $c$.~\cite{felser98} The three main FS sheets are a large hole-like sheet around $\Gamma$ (3D) and a pair of small electron-like and very flat sheets from quasi-one dimensional (q1D) bands.~\cite{felser98,stowe98,yokoya05} The latter are formed by Te~5$p_x$-electrons of the Te$^{(2)}$ and Te$^{(3)}$ sites being adjacent to the van-der-Waals gap of the crystal structure, while the 3D sheet has a mixed Zr~4$d$, Te~5$p$ character. The CDW gap, which persists to temperatures far above $T_{CDW}$ due to short-range ordered fluctuations, has been observed by angle-resolved photoemission spectroscopy (ARPES)~\cite{yokoya05} and infrared absorption spectroscopy.~\cite{perucchi2005} \zrte{} can be intercalated leading to bulk SC for both Ni ($T_c=3.1$ K)~\cite{lei11} and Cu ($T_c=3.8$ K).~\cite{zhu11} Coexistence of CDW and bulk SC has been reported for these intercalation compounds, for a solid solution replacing some of the Te with Se.~\cite{zhu2016}  The CDW resistivity anomaly is shifted to lower temperatures and with a reduced amplitude, while optical absorption spectra appear to show an increase of the CDW gap size in both Cu$_x$\ce{ZrTe3} and Ni$_x$\ce{ZrTe3}.~\cite{mirri2014}

This paper reports a combined theoretical and experimental study of Ni intercalation in ZrTe$_3$. Using {\em ab-initio} calculations we identify the energetically favourable intercalation sites and analyse the structural and electronic changes for two dopant concentrations. Samples at varying Ni concentration have been prepared and characterised as previously described.~\cite{lei11} Details of growth, characterisation and analysis of $T_{CDW}$ and $\Delta R/R$ reported in Fig.~\ref{Fig1} are described in Section~\ref{SectExp}. Momentum-resolved experiments by x-ray diffraction and ARPES confirm the unchanged presence of the CDW through the series, including the superconducting sample and reveal subtle changes to the underlying electronic structure that are compared to the {\em ab-initio} calculation results. The paper is structured as follows: Sect.~\ref{SectMethods} summarises the theoretical and experimental methods. Results and observations are described in Sect.~\ref{SectResults}. The implications of these results are discussed in Sect.~\ref{SectDiscussion}. The final section lists the conclusions that can be drawn from the study. Two appendices give additional results for the benefit of the interested reader.

\section{Methods}
\label{SectMethods}

\subsection{Theoretical}
\label{SectTheo}

All calculations were performed using the Vienna \textit{ab-initio} Simulation Package (VASP),\cite{Kresse1993,Kresse1994,Kresse1996a,Kresse1996} a periodic plane wave density functional theory code.
The Projector Augmented Wave (PAW) method was used in conjunction with pseudopotentials to describe the interactions between core and valence electrons.\cite{Kresse1999}
Convergence with respect to the plane wave basis set and \textit{k}-point sampling was performed, with a cut-off energy of \SI{350}{\electronvolt} and \textit{k}-point grid of $\Gamma$-centred $14 \times 10 \times 6$ found to be sufficient for the 8 atom unit cell of \ce{ZrTe3}.
This study employed the PBEsol functional,\cite{Perdew2008} a version of the Perdew Burke and Ernzerhof (PBE) functional revised for solids.\cite{Perdew1996}
PBEsol has been shown to accurately reproduce the structure parameters of layered systems held together by weakly dispersive interactions, without the need for an explicit van-der-Waals correction.\cite{Travis2016,Ganose2016,Biswas2017}
The structures were deemed converged when the sum of all forces on each atom totalled less than \SI{10}{\milli\electronvolt\per\angstrom}.
In order to accurately describe the electronic properties of \ce{ZrTe3}, spin--orbit coupling (SOC) effects were included for density of states and band structure calculations.\cite{Hobbs2000}
Bader charge analysis was performed using the bader code.\cite{Yu2011,Henkelman2006}

Defect calculations were performed in both a $3 \times 2 \times 2$ supercell (containing 96 atoms) and a $3 \times 4 \times 2$ supercell (192 atoms), using the PBEsol functional.
The formation energy, $\Delta H_f$, of a defect, $D$ was calculated as:
\begin{equation}
	\Delta H_f^{D}= (E^{D} - E^H) + \sum{\{n_i ( E_i + \mu_i )\}}
\end{equation}
where $E^{D}$ is the energy of the defected supercell and $E^H$ is the energy of the host supercell.
The second term represents the energy change due to losing an atom, $i$, to a chemical reservoir: $n_i$ is the number of atoms of each type lost, $E_i$ is the elemental reference energy calculated from the element in its standard state, and $\mu_i$ is the chemical potential of the atom. The formation energies of all competing phases are reported Appendix~\ref{labelAppendixFormation}.

Supercell calculations, due to shrinking of the Brillouin zone, result in folded band structures which cannot be easily compared to the bulk band structure.
To circumvent this issue, primitive cell representations of supercell band structures were obtained using the band unfolding code BandUp,\cite{Medeiros2014, Medeiros2015} based on the methodology described by Popescu and Zunger.\cite{Popescu2012}

\subsection{Experimental}
\label{SectExp}

Single-crystal Ni$_x$ZrTe$_3$ have been grown via the chemical vapour transport method as described in Ref.~[\onlinecite{lei11}] in four batches  with varying Ni content $x$. Electrical resistivity measurements (Fig.~\ref{Fig1}) show that the CDW transition - defined as the midpoint between the onset $T_{CDW1}$ and completion $T_{CDW2}$ of the resistivity anomaly - becomes rather narrow and is even somewhat enhanced in sample A. Samples B and C suggest gradual suppression of the CDW transition as well as the onset of  superconductivity in sample C. The fourth sample SC (previously described) is superconducting below $T_c = 3.1$~K at $x = 0.052\pm0.003$~\cite{lei11}.  The Ni content is increasing from sample A to sample SC. The precise Ni content in samples A-C was not determined explicitly as it is below the detection limit of about $x_{lim} = 0.04$ in our EDAX and XPS characterisation methods. We can assume the following Ni content of the samples: A: Ni$_{0.01}$\zrte, B: Ni$_{0.02}$\zrte{} and C: Ni$_{0.03}$\zrte. 

Experiments of ARPES were performed at the HR-ARPES instrument at beamlline I05
at Diamond Light Source.~\cite{hoesch17} The samples were cleaved at the cryogenic measurement temperature of $T=7$~K in an ultrahigh vacuum better than $1.2\times10^{-10}$~mbar. Electron spectra were measured with the photon polarisation held in the sample ($b^*-c^*$)-plane ($p$-polarised geometry) and  FS maps were acquired by rotating the sample about the $a$-axis. The photon energy was set to $h\nu=70$~eV in most data sets resulting in an energy resolution of 18~meV.

Diffuse x-ray scattering experiments were performed at the EH2 instrument at beamline I19 at Diamond Light Source at a wavelength of $\lambda = 0.6889$~\AA. The sample were held at $T=20$~K and exposed to the monochromatic x-ray beam during step-by-step rotations with synchronised acquisitions by the CCD x-ray detector. The detector was deliberately over-exposed in the strong main-lattice diffraction peaks to bring out the much weaker superstructure peaks due to the periodic lattice distortion (PLD) well above the noise background. The data sets were indexed using Crysalis software,~\cite{crysalis06} yielding crystal parameters with large error margins due to the overexposure. Full high-symmetry planes of scattering have been reconstructed from the data as shown in  Fig.~\ref{XRD}. The reciprocal space position of incommensurate superstructure reflections was determined as the centre of gravity of the intensity in as small area contained in the circles marked in Fig.~\ref{XRD}. These positions as well as the unit cell volume, which is more reliably determined than the individual lattice parameters are summarised in Table~\ref{XRDsummary}. 

\section{Results}
\label{SectResults}

\subsection{Calculated properties of bulk \ce{ZrTe3}}

\begin{figure}[h!]
\centering
\includegraphics[width=0.48\textwidth]{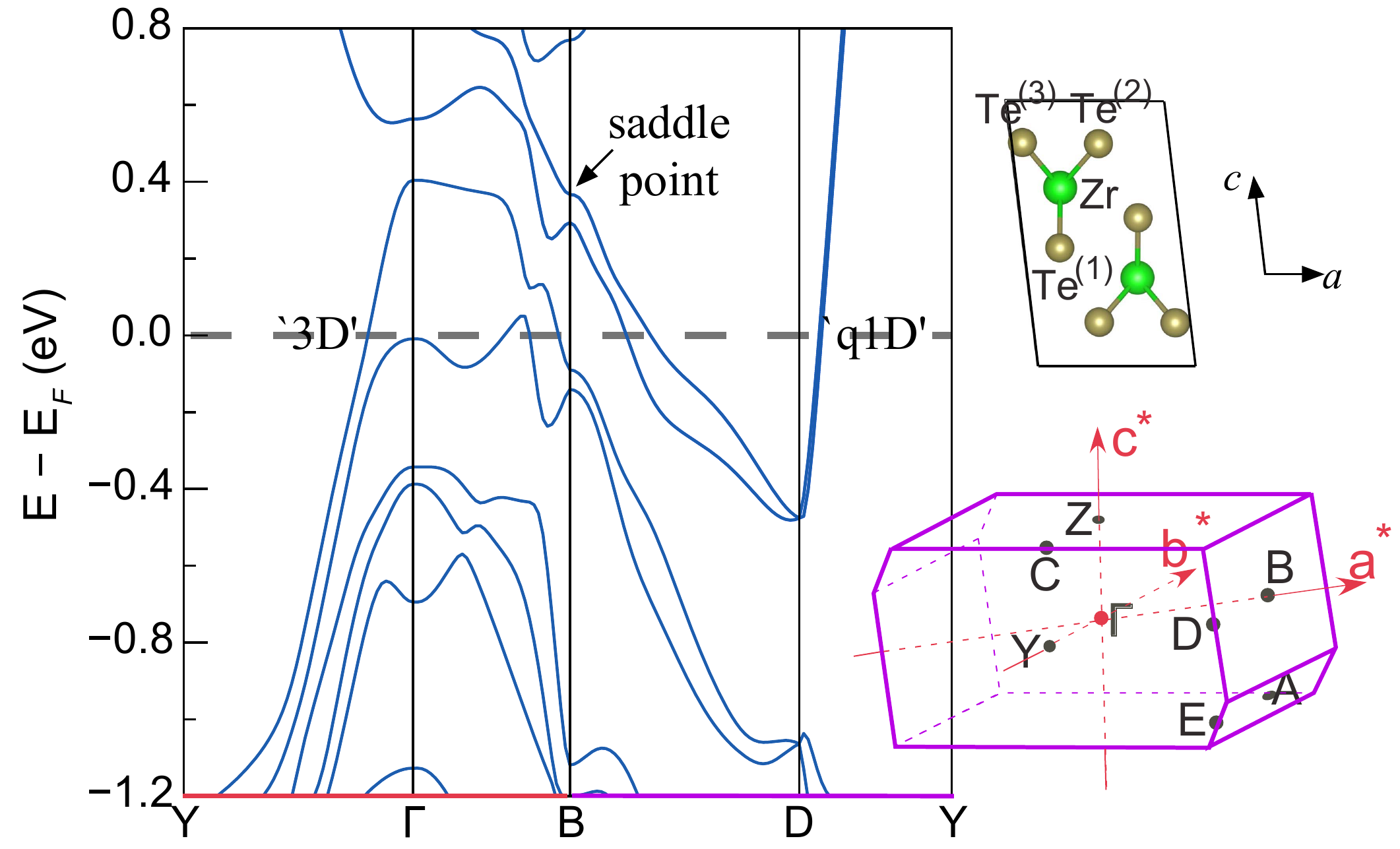}
\caption{Calculated band structure and converged unit cell of \ce{ZrTe3}.}
\label{fig:electronic-structure} 
\end{figure}

We first turn our attention to the predictions for pure \ce{ZrTe3}. 
The PBEsol optimised lattice parameters of \ce{ZrTe3} are given in Table \ref{tab:lattice-constants}.
The lattice parameters show good agreement with experiment,\cite{Seshadri1998} aside from a slight underestimation of the \textit{b} parameter by \SI{1.4}{\percent}.
The band structure and converged unit cell of \ce{ZrTe3} are shown in Figure \ref{fig:electronic-structure}, with the Brillouin zone for the $P 2_1 /m$ space group also provided.
PBEsol reproduces the metallic nature of \ce{ZrTe3} with the correct number of Fermi crossings along Y-$\Gamma$ (one, marked `3D') and D-Y (two degenerate, marked `q1D'). Slight discrepancies to the ARPES observation exist, including the position of the  saddle point of the q1D band at B [$E-E_F=0.38$, compared to $E\simeq E_F$ in experiment (van-Hove singularity)~\cite{yokoya05}]. In the following, all features will be investigated for their trends as a function of doping and are easily identified through their dispersions.

\begin{table}[h]
\centering
\caption{Lattice parameters and angles for \ce{ZrTe3} relaxed using PBEsol compared with experiment.\cite{Seshadri1998} $\alpha$ and $\gamma$ were both found to be \SI{90}{\degree}.}
\label{tab:lattice-constants}
\begin{tabular}{@{}lccccc@{}}
\toprule
                & \textit{a} (\r{AA}) \; & \textit{b} (\r{AA}) \; & \textit{c} (\r{AA})\;  & $\beta$ ($\deg$)\;  & vol. ($\mathrm{\AA}^3$)\;  \\ \midrule
\; Experiment\cite{Seshadri1998}     \;  \;         & 5.873   \;   & 3.908   \;     & 10.054  \;    & 97.85  \; & 228.6                \\
\; PBEsol    \;  \;      & 5.847   \;    & 3.854 \;      & 10.100  \;    & 97.02     \; & 225.9              \\
\; Difference (\%)\;  \;  & -0.43  \;     & -1.40  \;     & 0.47   \;     & -0.85      \; & -1.2             \\ \bottomrule
\end{tabular}
\end{table}

\subsection{Ni Defects}

To test the effect the of Ni incorporation on the electronic and geometric structure of \ce{ZrTe3}, we have calculated the energetics of a range of Ni interstitial and substitutional defects.
Three interstitial ($\mathrm{Ni}_i$) sites were identified within the van der Waals gap, comprising two pentacoordinated ($\mathrm{Ni}_i^1$ and $\mathrm{Ni}_i^2$) and one tetrahedral ($\mathrm{Ni}_i^3$) sites.
Additionally, a substitutional defect, $\mathrm{Ni_{Zr}}$, was trialled.
Table \ref{tab:defect-formation-energies} shows the defect formation energies of the defects for both supercell sizes considered.
A single interstitial in the $3 \times 4 \times 2$ and $3 \times 2 \times 2$ supercells corresponds to doping concentrations $x_{3 \times 4 \times 2}=2.1 \%$ and $x_{3 \times 2 \times 2}=4.2 \%$, respectively.

\begin{table}[h]
\centering
\caption{Defect formation energies, $\Delta H_f^{D}$, for a range of interstitial and substitutional defects, calculated within $3 \times 4 \times 2$ and $3 \times 2 \times 2$ supercells using the PBEsol functional.}
\label{tab:defect-formation-energies}
\begin{tabular}{@{}lcc@{}}
\toprule
                   & \multicolumn{2}{c}{$\Delta H_f^{D}$ (eV)}     \\ \cmidrule(l){2-3} 
Defect    \; \;          & \; $3 \times 4 \times 2$\; \;  & \; \; $3 \times 2 \times 2$ \; \; \\ \midrule
$\mathrm{Ni}_i^1$  & 0.594                 & 0.594                 \\
$\mathrm{Ni}_i^2$  & 0.594                 & 0.594                 \\
$\mathrm{Ni}_i^3$  & 1.462                 & 1.516                 \\
$\mathrm{Ni_{Zr}}$ & 1.743                 & 2.160                 \\ \bottomrule
\end{tabular}
\end{table}

\begin{figure}[htb]
\centering
\includegraphics[width=0.48\textwidth]{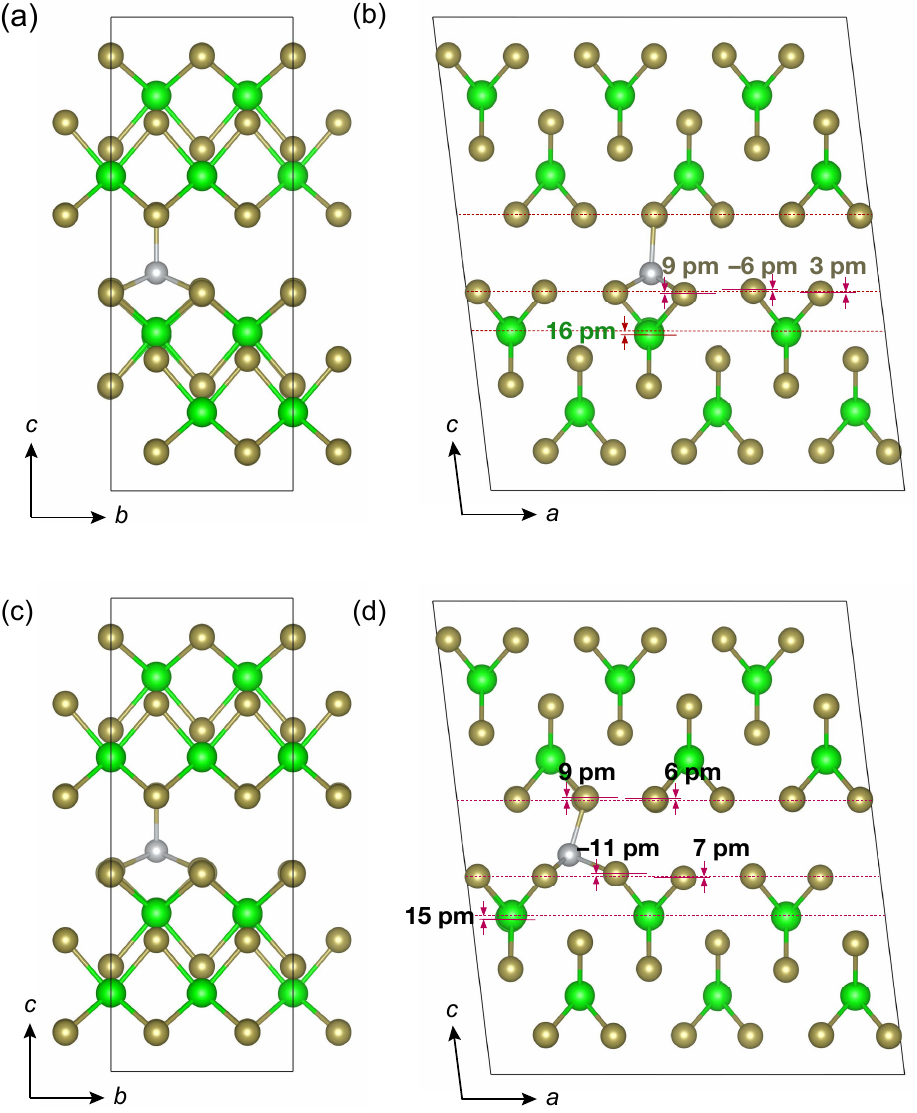}
\caption{Converged $3 \times 2 \times 2$ supercell of \ce{ZrTe3} containing a $\mathrm{Ni}_i^1$ defect (a, b) and  a $\mathrm{Ni}_i^3$ defect (c, d). Zr, Te, and Ni atoms indicated by green, gold, and silver spheres, respectively. Displacements of Zr and Te due to the presence of Ni are shown in (b) and (d). Positive values correspond to displacements away from the van-der-Waals gap.}
\label{fig:defect-sites} 
\end{figure}

The formation energy of the $\mathrm{Ni_{Zr}}$ defect is significantly greater than that of the interstitial defects, indicating that the substitutional defect is unlikely to be present in large concentrations.
For the interstitial defects, both pentacoordinated defects relaxed to the same position [Fig.~\ref{fig:defect-sites}(a,b)], and accordingly share the same relatively low formation energy ($\sim$\SI{0.6}{\electronvolt}).
The tetrahedrally-coordinated interstitial defect also relaxed to a pentacoordinated site [Fig.~\ref{fig:defect-sites}(c,d)], but showed a much larger formation energy ($\sim$\SI{1.5}{\electronvolt}).
The primary difference between the two interstitial sites can be seen in Figures \ref{fig:defect-sites}b and d, with the low formation energy $\mathrm{Ni}_i^1$ defect situated directly above a lattice Zr site [Fig.~\ref{fig:defect-sites}(b)], in contrast to the higher-energy $\mathrm{Ni}_i^3$, which defect occupies the space adjacent to a link between \ce{(ZrTe3)}$_\infty$ chains  [Fig.~\ref{fig:defect-sites}(d)].
We note that the defect formation energies show qualitative agreement across both doping concentrations, with only the $\mathrm{Ni_{Zr}}$ defect showing a significant dependence on supercell size.

An analysis of the atomic displacements in the presence of fully converged $\mathrm{Ni}_i^1$ and $\mathrm{Ni}_i^3$ defects is shown in Fig.~\ref{fig:defect-sites}. This further confirms the larger energetic cost of the  $\mathrm{Ni}_i^3$ defect [Fig.~\ref{fig:defect-sites}(c,d)] as  Zr-Te bonds on either side of the van-der-Waals gap are distorted. We concentrate on the energetically favourable $\mathrm{Ni}_i^1$ defect [Fig.~\ref{fig:defect-sites}(a,b)]. Remarkably, the largest displacement (16~pm) is found for the Zr atom close to the Ni site, while the Te atoms forming short bonds with Ni are displaced by less than 10 pm. This displacement leads to a buckling of the Te$^{(2)}$-Te$^{(3)}$ chains ({\em c.f.} Fig.~\ref{fig:electronic-structure}) hosting the q1D band with the Te$^{(2)}$ adjacent to the defect displaced by $+9$~pm away from the van-der-Waals gap. The buckling even extends to the neighbouring (\ce{ZrTe3})$_\infty$ chain ($-6$~pm and $+3$~pm displacement of Te$^{(3)}$ and Te$^{(2)}$, respectively).

For both doping concentrations, the change in lattice parameters is small (less than \SI{-0.35}{\percent}), as expected based on the relatively small average atomic displacements discussed above.  Overall chemical pressure effects are therefore not expected to play a major role and the changes observed in the electronic structure stem from local effects.

The charge transfers arising from the $\mathrm{Ni}_i^1$ defect were determined by a Bader analysis for the \SI{4.2}{\percent} doped system. The Ni atom possesses a slight negative charge of \SI{0.07}{\elementarycharge}, almost equivalent to neutral.
The Zr and Te atoms in the vicinity of the Ni site correspondingly retain charges close to the values found in the undoped system. 

\subsection{Band Unfolding}

\begin{figure*}[htb]
\centering
\includegraphics[width=0.99\textwidth]{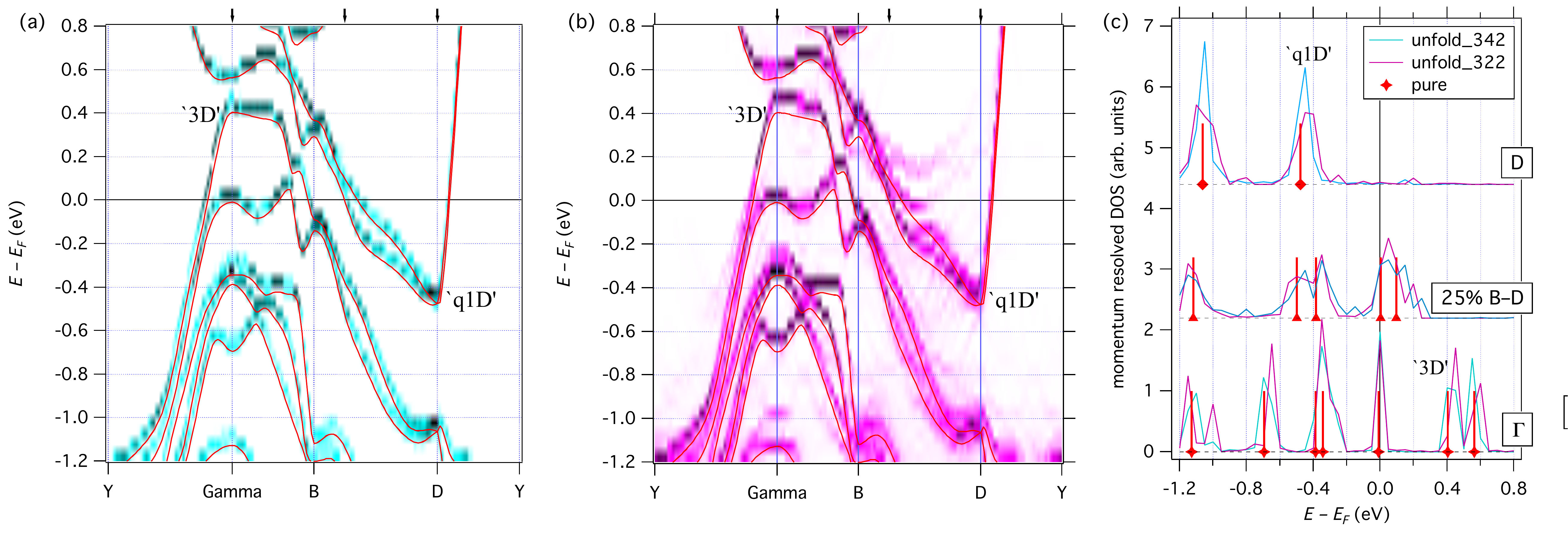}
\caption{Unfolded band structure of \ce{ZrTe3} for the supercells with Ni concentrations of \SI{2.1}{\percent} (a) and \SI{4.2}{\percent} (b). Panel (c) shows the extracted momentum-resolved density of states at three momentum positions marked by arrows. Overlaid as sharp curves are the band dispersions of \ce{ZrTe3}. All data are calculated using PBEsol+SOC.}
\label{fig:defect-electronic-structure} 
\end{figure*}

To compare the momentum-resolved density of states of the supercell calculations to pure \zrte{} and the experiment the unfolded band structures are shown in Figure \ref{fig:defect-electronic-structure}. In this way, the perturbations to the primitive cell band structure, due to the introduction of a $\mathrm{Ni}_i^1$ defect, can be visualised. 
In the \SI{2.1}{\percent} doped system, only slight changes to the bands around the Fermi level are observed, with the bands along B -- D showing the greatest degree of modification. 
Indeed, comparison with the overlaid band structure of pure \ce{ZrTe3}, indicates that the defected system retains a ``bulk-like'' electronic structure.
The perturbations to the bands along B -- D become more apparent in the \SI{4.2}{\percent} doped system, as evidenced by the greater spectral weight of the bands seen $\sim$\SI{0.2}{\electronvolt} above the Fermi level.
Between D -- Y, the q1D band dispersion remains unchanged with the binding energy shifted by +0.026~eV and +0.051~eV for \SI{2.1}{\percent} and \SI{4.2}{\percent}, respectively. Note that the band-folding around the middle along B -- D seen in particular in Fig.~\ref{fig:electronic-structure}(b) is an artefact due to the doubling of the supercell in this direction.

\subsection{Angle-resolved photoemission spectroscopy (ARPES)}

\begin{figure*}
\centerline{\includegraphics[width = .34\textwidth]{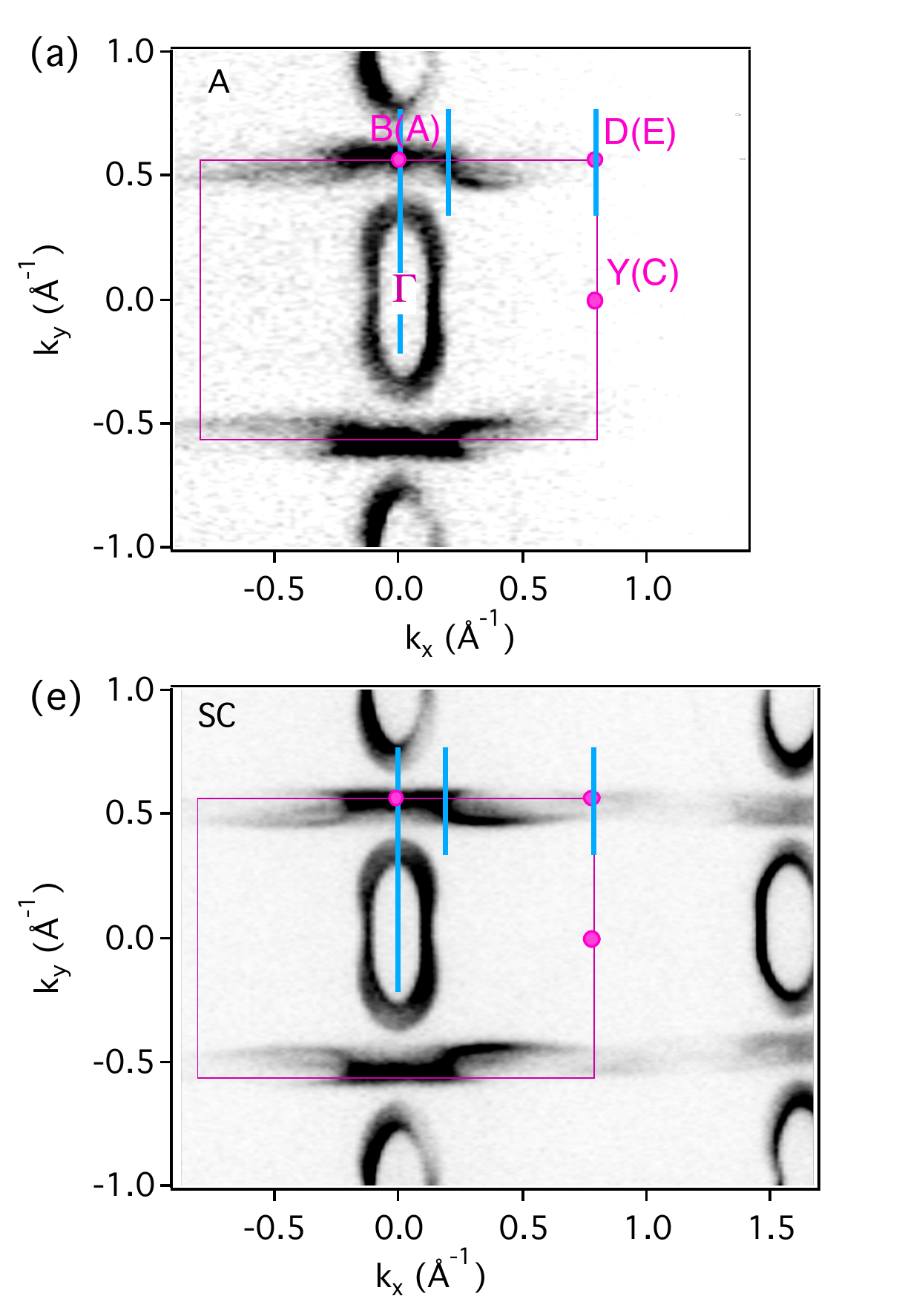}\includegraphics[width = .56\textwidth]{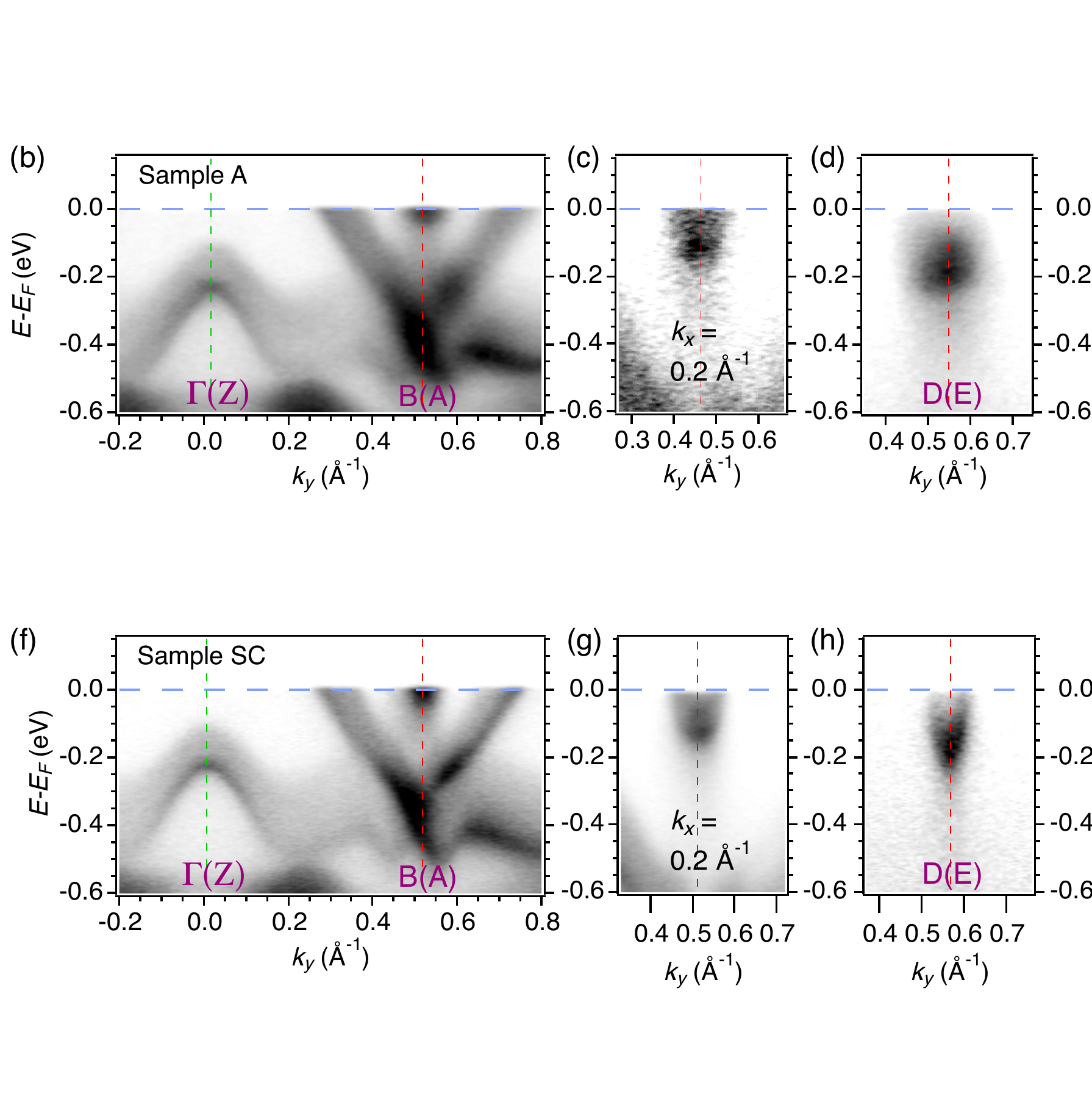}}
\caption{Fermi surface maps of samples A (a) and SC (e) acquired by ARPES at a temperature $T = 7$~K. Band dispersion maps along lines of constant $k_x$ indicated in (a,e) are shown for sample A (b-d) and SC (e-g).}
\label{FigARPES}
\end{figure*}

The electronic structure as observed by ARPES is shown in Fig.~\ref{FigARPES} for samples A and SC. Samples B and C have given data of similar quality. For sample A, the Fermi surface mapping data [panel (a)] and thus also the observation of band dispersions [panels (b,c)] are more blurry, which relates to sample alignment to a less favourable measurement spot. Data from sample SC are of exceptional quality, namely the dispersion of the q1D band near \dbar\ is seen as a sharp parabola shape. We note that no CDW band backfolding is visible, not even as a broadening of features close to $E_F$. Within the momentum resolution thus given by the quality of sample orientation (mosaic) at the measurement spot, there are no obvious changes of Fermi wave vectors or additional/missing FS pockets in any of the samples. The characteristic depletion of spectral weight due to CDW formation in the q1D band close the the Brillouin zone boundary along \bbar\ -- \dbar\ is visible as a missing intensity in all Fermi surface maps. The spectral weight is removed in a region from about $k_x = 0.35 \ \invA$ to \dbar{}  at $k_x=0.8 \ \invA$.

\begin{figure*}[th!]
\centerline{\includegraphics[width = .92\textwidth]{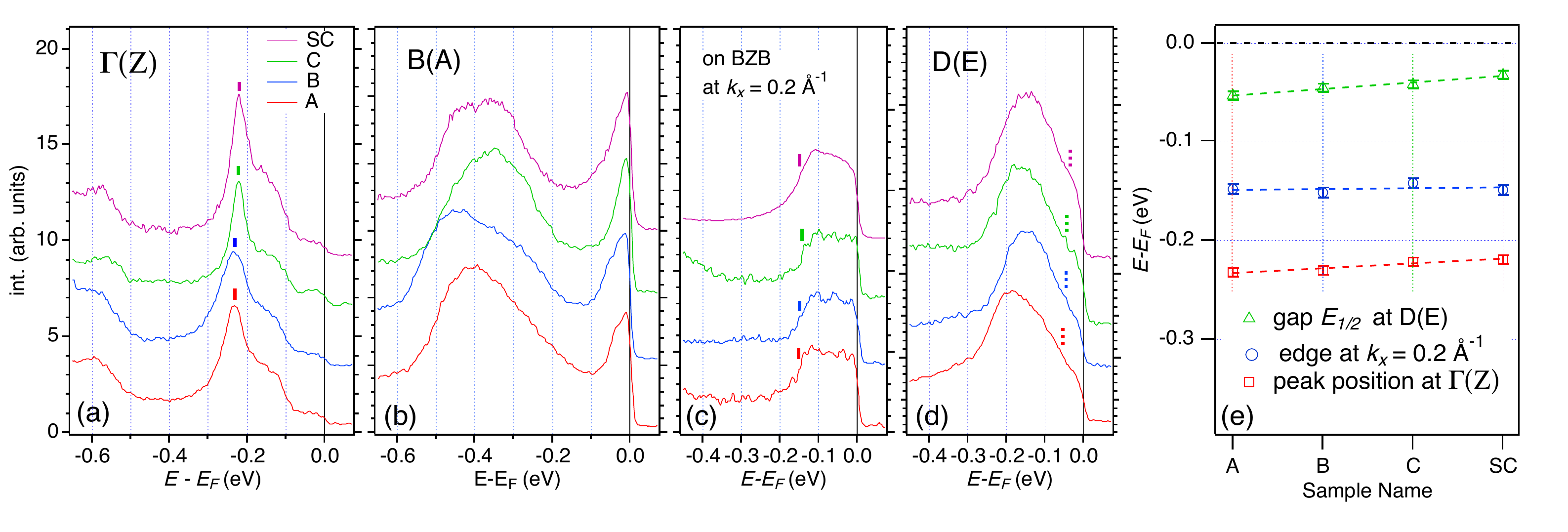}}
\caption{(a-d) Spectra at specific high symmetry points as indicated at the top of each panel for all four samples at cuts at $T = 7$~K. (e) Extracted peak positions from panel (a) and lower edge positions from panel (d) as a function of sample name with increasing Ni content.}
\label{FigBE_trends}
\end{figure*}

Energy distribution curves (spectra as a function of $E-E_F$ at specific spots in momentum space) are shown in Fig.~\ref{FigBE_trends}(a-c). Suitable momentum-integrations over small regions have been applied to equalise the varied momentum resolution between samples. While some variation of spectral shapes from sample to sample are observed, the key features remain unchanged, namely sharp peaks at $E-E_F=-0.23$~eV at \gammabar\ and at $E-E_F=-0.02$~eV at \bbar\ [panels (a,b)] and a band of intensity up to $E=E_F$ at $k_x = 0.2\ \invA$ and at \dbar{} [panels (c,d)]. 

These data are inspected for trends of binding energies across the series in Fig.~\ref{FigBE_trends}(e). First concentrating on the \dbar{} point, we analyse the position of half height intensity $E_{1/2}$ between the peak at $E-E_F= -0.16$~eV and the background  ($E>E_F$). This half-height  serves us as a measure of the CDW gap size. Caution is, however, required in its interpretation as a change may be due to a change of the underlying band structure as well as due to the magnitude of spectral weight depletion. A trend of decreasing gap size from $E_{1/2}-E_F=-0.054$~eV in sample A to $E_{1/2}-E_F=-0.033$~eV in sample SC is observed [top trace in Fig.~\ref{FigBE_trends}(e)]. The same q1D band is seen also at $k_x = 0.2\ \invA$, where no temperature-dependent spectral weight depletion is occurring. We determine the high energy onset of spectral weight around $E-E_F = -0.15$~eV for each sample. The data show a very high consistency and no changes of this onset are found within the error bars ($\pm0.01$~eV) across the series [middle trace in Fig.~\ref{FigBE_trends}(e)]. Similarly at \bbar, the sharp peak due to a van-Hove singularity at $E-E_F =-0.01$~eV~\cite{yokoya05} is seen unchanged across the series, thus indicating that no Ni-dependent charge transfer is observable on the Te-derived q1D. Finally the peak at \gammabar\  shows a decreasing binding energy (increasing $E$) from  $E-E_F=-0.234(6)$~eV in sample A to  $E-E_F=-0.222(6)$~eV in sample SC [lowest trace in Fig.~\ref{FigBE_trends}(e)]. This peak derives from the apex of a down-dispersing band, also seen in the calculations, that is identified as almost purely Zr 4$d$ derived. 

\subsection{X-ray Diffraction}

\begin{figure}
\vspace{4mm}
\centerline{\includegraphics[width = .45\textwidth]{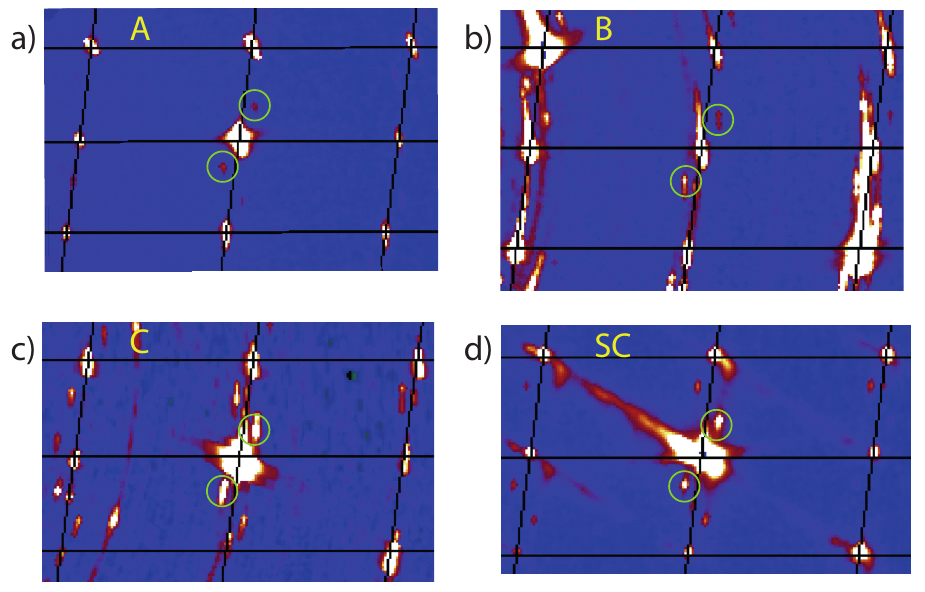}}
\caption{Diffraction plane ($h0l$) of samples A, B, C, and SC spanning from (300) in the top left to (50$\overline{2}$) in the bottom right corner of each panel. Blue (yellow) corresponds to low (high) intensity and the strong main lattice spots are over-exposed. The measurement temperature was $T=20$~K. The position of the superstructure spots due to the PLD  is marked with green circles.}
\label{XRD}
\end{figure}

Selected data from a small section of the ($h0l$) plane around $\vec{Q}$=\fourbar{} are shown in Fig.~\ref{XRD}. The over-exposure leads to large and blurry appearance of the main-lattice diffraction spots. The low intensity tail of these spots with thermal diffuse scattering character is contained in these spots. In addition some spurious features and arcs  centred on $\vec{Q}=0$ are seen, resulting from small misoriented parts of the sample that can be ignored in the analysis. Systematically observed superstructure spots compatible with $\vec{q}_{CDW}$ are present in all four samples. Namely at $\vec{Q}=(40\overline{1})\pm \vec{q}_{CDW}$ (centre of each panel in Fig.~\ref{XRD}) the PLD is clear and sharp.

\begin{table}[h]
\caption{PLD superstructure modulation vector and unit cell volume at $T=20$~K determined from x-ray diffraction data.}
\begin{center}
\begin{tabular}{ccc}
\hline
Sample       &       PLD                       &   Volume (\AA$^3$)\\
\hline
pure (lit)&(0.071(5),\ 0,\ 0.336(10) \cite{eaglesham84}& 228.6\cite{Seshadri1998}\\
A		  &   (0.074(3),\ 0,\ 0.31(2))  &   231.3(2)    \\
B                &    (0.072(2),\ 0,\ 0.34(2))  &   231.8(1)     \\
C                &    (0.071(2),\ 0,\ 0.34(2))  &   231.6(2)    \\
SC             &    (0.072(2),\ 0,\ 0.33(2))  &   232.0(2)     \\
\hline
\end{tabular}
\end{center}
\label{XRDsummary}
\end{table}%

Table~\ref{XRDsummary} summarises the observed PLD modulation vector $\vec{q}_{CDW}$ and the unit cell volume for each sample.  The latter shows a slight increasing trend, compatible with the previously reported increase of $a$ and $c$-axis lattice parameters.\cite{lei11,mirri2014} $\vec{q}_{CDW}$ as well as the strength of the superstructure diffraction  peaks compared to the main lattice peaks is  unchanged within the experimental uncertainty. In the highest doped sample SC, a pattern of diffuse scattering planes, visible as a diagonal lines of intensity {\em e.g.} connecting (300) to \fourbar. This diffuse scattering pattern is independent of temperature up to room temperature and may be dislocation-related~\cite{furuseth75} and could correspond to temperature-independent diffuse scattering features seen in electron diffraction from pure \ce{ZrTe3}.\cite{eaglesham84}

\section{Discussion}
\label{SectDiscussion}

With its multi-band Fermi surface, \zrte{} can accommodate a coexistence of CDW and SC by offering DOS at $E_F$ for superconductivity on one sheet while another sheet is gapped by the CDW. The experiments we present in this study are spatially averaging and  a simultaneous observation can also arise from in-homogeneity effects. We observe the CDW consistently in all samples and in all three probes: in resistivity (Fig.~\ref{Fig1}) the CDW anomaly is first  enhanced in magnitude $\Delta R/R_1$ and then reduced in both magnitude and $T_{CDW}$ as well as broadened with increasing Ni content. The XRD shows an unchanged $\vec{q}_{CDW}$ and the PLD remains clearly visible with nearly unchanged intensity up to the highest doped sample SC (Table~\ref{XRDsummary}). The ARPES data show the full complexity of the CDW gapping, which occurs only on part of the q1D FS sheets as previously reported.~\cite{yokoya05} Again, the gapping is unchanged in momentum-space extent and only a slight reduction in $E_{1/2}$ at \dbar{} (Fig.~\ref{FigBE_trends}) may hint at a reduction in CDW. It should be noted that ARPES is sensitive to local effects, thus it will show the fluctuating as well as long-range ordered CDW while XRD shows the long range order only. We find thus that the CDW persists up to the highest observed doping, where bulk SC with $T_c = 3.1$~K is been observed.~\cite{lei11} At least a fluctuating CDW is thus expected to persist to even higher doping and coexistence of CDW and SC is strongly supported by our data. 

The modifications of the electronic structure that are observed both in ARPES and in the calculations are smaller than the observations reported in transition-metal-intercalated \ce{TiSe2}.~\cite{cui06,zhao07,Jishi08} Both the calculations as well as the ARPES data show a virtually unchanged q1D band close to \dbar, although a small change of binding energy ($<0.05$~eV) may exist. Also along the BZB, closer to \bbar{} where no CDW gapping is observed, the ARPES data show no shift of binding energy $E$ and only a small upward shift of $E$ (0.05 eV) is observed in the calculations. A shift of similar size (0.015~eV) is observed in the ARPES data at \gammabar{} for the Ni 4$d$-derived band at $E-E_F =-0.23$~eV. This shift is compatible with the charge transfer from Zr to Ni.  The stability of the CDW can be affected by the filling of the q1D band, which is seen reduced in the calculations, compensated by a similar decrease in hole carrier concentration in the 3D band. These changes are subtle and smaller than the ones that were identified as unbalancing the CDW in \zrte{} under pressure.~\cite{hoesch2016} More important therefore than an unbalancing of carriers in the FS sheets are the effects of disorder from the Ni intercalation.

The dome-shaped evolution of $\Delta R/R_1$ (Fig.~\ref{Fig1}) and the rapid suppression of $T_{CDW}$ from samples B to SC is reminiscent of the dome of CDW under hydrostatic pressure, which is first enhanced and then suppressed rapidly.\cite{yomo05} The magnitude of the enhancement of $T_{CDW}$ is, however, much larger with pressure (nearly 40~K). Our data show furthermore that the modulation $\vec{q}_{CDW}$ is not changing with Ni content, while a rotation and change of size of  $\vec{q}_{CDW}$ was observed under hydrostatic pressure.\cite{hoesch2016} The small change of lattice constant ($\sim +0.3 \%$ of unit cell volume change), compared to $-8 \%$ up to 5~GPa pressure, is indicative that chemical pressure plays a minor role in the quenching of the CDW. Instead the local buckling of the Te sheet close to the van-der-Waals gap, adjacent to the Ni defect [Fig.\ref{fig:defect-sites}(b)] is likely to have a strong effect on the long range order of the CDW by pinning. The displacement pattern of the long wavelength (small $q$) CDW is unknown, though likely to be transverse to the propagation direction.\cite{hoesch09ixs} The fact that the buckling extends to neighbouring ZrTe$_3$ chains may be due to a locally established CDW-like distortion. The Ni defects can thus locally enhance a distortion, similar to the  CDW displacement, which  suppresses the  the long-range order due to the stochastic positioning of the defects. 

\section{Conclusions}

The combined experimental and theoretical study presented here allows the following conclusions to be drawn. The effects of Ni intercalation are mostly local, in the close vicinity of the Ni site. Changes in unit cell parameters are small and therefore an overall chemical pressure effect plays a minor role. This is also supported by the evolution of $\vec{q}_{CDW}$, which remains constant rather than changing as under hydrostatic pressure.~\cite{hoesch2016} The Ni site is unambiguously identified as an intercalation site in the van-der-Waals gap with a five-fold coordinated site directly above a Zr atom being energetically favourable over a similar site above a link between (\ce{ZrTe3})$_\infty$ chains. The key effects close to the Ni intercalant site is a buckling of the Te sheet adjacent to the van-der-Waals gap and a shift of the Zr position, even larger than the amplitude of the buckling. The charge transfer from the (\zrte{})$_\infty$ chain to Ni is negligible. This buckling may correspond to a displacement pattern similar to the CDW and and acts as pinning centre, which quenches the long-range order. Within the ability of the experiment to determine this value, the CDW gap is also unchanged on Ni intercalation, both in extent in momentum space and in size ($E_{1/2}$). The latter is, if anything, slightly reduced in the highest intercalated sample SC.

\acknowledgments
We wish to thank P.D.C. King for help with preliminary experiments and T. K. Kim, M. Warren and D. Allan for help with the experiments and fruitful discussions. We are grateful to F. Baumberger for use of his ARPES data analysis software and to Diamond Light Source, where access to beam lines I19 (MT8776) and I05 (NT4924 and NT12226) contributed to the results presented here. One of us (MH) would like to acknowledge use of the Bodleian Libraries (Oxford) at the time of writing the manuscript. Work at Brookhaven National Laboratory was supported by the U.S. Department of Energy, Office of Science, Office of Basic Energy Sciences, under Contract No. DE-SC0012704.

\appendix

\section{Formation of \ce{ZrTe3}}
\label{labelAppendixFormation}

Through varying the chemical potentials, we can simulate the effect of experimentally varying the partial pressures in the formation of \ce{ZrTe3}.
These potentials are defined within the global constraint of the calculated enthalpy of the host, in this case: \ce{ZrTe3}: $\mu_\mathrm{Zr} + 3\mu_\mathrm{Te} = \Delta H_f^\mathrm{ZrTe3}$. 
To avoid precipitation into solid elemental Zr and Te, we also require $\mu_\mathrm{Zr} \leq 0$ and $\mu_\mathrm{Te} \leq 0$. 
Lastly, the chemical potentials are further constrained in order to avoid decomposition into a range of competing binary compounds. 
For \ce{ZrTe3}, the competing phases considered were \ce{Zr3Te}, \ce{Zr5Te4}, \ce{Zr5Te6}, \ce{ZrTe}, \ce{ZrTe3}, and \ce{ZrTe5}. 
The formation energies of all competing phases, calculated using PBEsol, are provided in Table \ref{tab:competing-phases}.

\begin{table}[h]
\centering
\caption{Full list of competing phases considered when calculating the chemical potential space of \ce{ZrTe3}, along with their corresponding formation energies}
\label{tab:competing-phases}
\begin{tabular}{@{}lc@{}}
\toprule
\; Compound \; \;& $\Delta H_f$ (eV) \; \\ \midrule
\; Zr3Te  \; \;  & -2.296     \\
\; Zr5Te4 \; \;   & -8.286     \\
\; Zr5Te6 \; \;  & -10.541     \\
\; ZrTe   \; \;  & -1.993     \\
\; ZrTe3  \; \;  & -2.910     \\
\; ZrTe5  \; \;  & -3.007     \\ \bottomrule
\end{tabular}
\end{table}

\section{Density of States}
\label{labelAppendixDOS}

\begin{figure*}
\includegraphics[width = 0.92\textwidth]{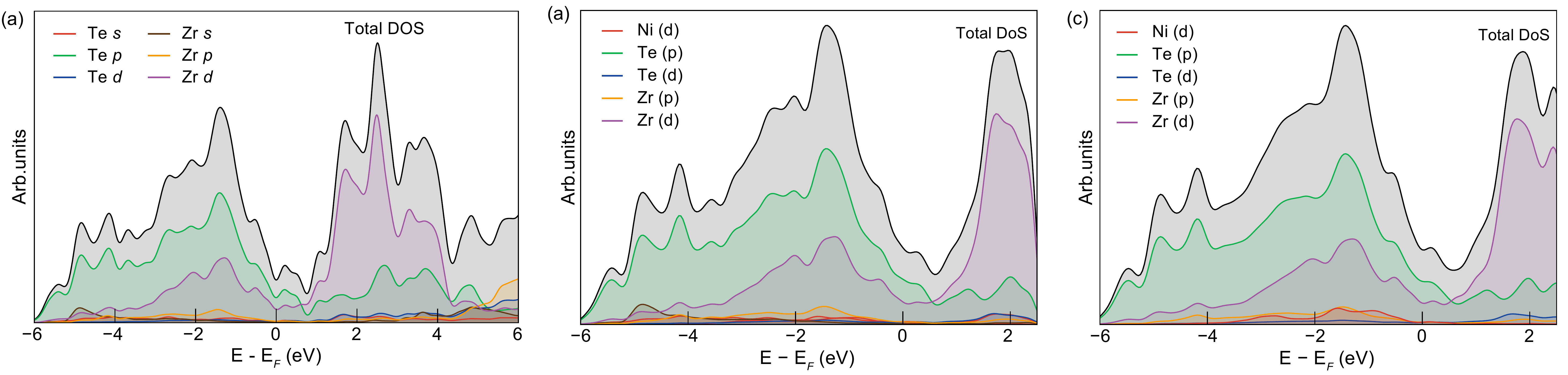}
\caption{Density of states for \zrte{} (a) as well as for the supercells with Ni concentrations of \SI{2.1}{\percent} (b) and \SI{4.2}{\percent} (c).}
\label{FigDOS}
\end{figure*}

The density of states of the $\mathrm{Ni}_i^1$ defect for both doping concentrations, calculated using PBEsol+SOC, are shown in Figures \ref{FigDOS}. The main difference to the bulk DOS is the introduction of the Ni 3\textit{d} states just below the Fermi level, between \num{-2} to \SI{0}{\electronvolt}, with only a slight contribution seen above the Fermi level. As expected, the density of states differ little across the doping concentrations, aside from a greater concentration of Ni states in the \SI{4.2}{\percent} doped system.

\bibliography{doping}

\end{document}